\documentclass[11pt, reqno]{amsart}
\usepackage{amsfonts}
\usepackage{mathrsfs}
\usepackage{amsmath}
\usepackage{epstopdf}
\usepackage{graphicx}
\usepackage{caption}
\usepackage{amssymb,amsfonts}
\usepackage{hyperref}
\hypersetup{hypertex=true,colorlinks=true,linkcolor=blue,anchorcolor=blue,citecolor=blue}

\newtheorem{thm}{Theorem}[section]
\newtheorem{lem}{Lemma}[section]

\newtheorem{prop}{Proposition}[section]
\numberwithin{equation}{section}

\def\pf{{\textit {Proof:} }}

\newcommand{\mysection}[1]{\section{#1}\setcounter{equation}{0}}

\newcommand{\bal}{\begin{aligned}}      \newcommand{\eal}{\end{aligned}}
\newcommand{\ba}{\begin{array}}      \newcommand{\ea}{\end{array}}
\newcommand{\bc}{\begin{center}}     \newcommand{\ec}{\end{center}}
\newcommand{\be}{\begin{enumerate}}  \newcommand{\ee}{\end{enumerate}}
\newcommand{\beq}{\begin{eqnarray}}  \newcommand{\eeq}{\end{eqnarray}}
\newcommand{\beQ}{\begin{eqnarray*}} \newcommand{\eeQ}{\end{eqnarray*}}
\newcommand{\bi}{\begin{itemize}}    \newcommand{\ei}{\end{itemize}}
\newcommand{\bt}{\begin{tabular}}    \newcommand{\et}{\end{tabular}}
\newcommand{\bdm}{\begin{displaymath}} \newcommand{\edm}{\end{displaymath}}

\newcommand{\Lrw}{\Longrightarrow}
\newcommand{\Llrw}{\Longleftrightarrow}

\def\qed{\hfill{Q.E.D.}\smallskip}

\newcommand{\ls}{\setlength{\baselineskip}{12pt}
	\setlength{\parskip}{3mm}}

\begin{document}
	
	\title[Geodesics]{Geodesics on metrics of Eguchi-Hanson type}
	
	\author[Y Yang]{Yekun Yang$^{\dag}$}
	\address[]{$^{\dag}$Schoal of Mathematics and Information Science, Guangxi University, Nanning, Guangxi 530004, PR China}
	\email{yangyk@st.gxu.edu.cn}
	
	\author[X Zhang]{Xiao Zhang$^{\flat}$}
	\address[]{$^{\flat}$Guangxi Center for Mathematical Research, Guangxi University, Nanning, Guangxi 530004, PR China}
	\address[]{$^{\flat}$Academy of Mathematics and Systems Science, Chinese Academy of Sciences, Beijing 100190, PR China}
	\address[]{$^{\flat}$School of Mathematical Sciences, University of Chinese Academy of Sciences, Beijing 100049, PR China}
	\email{xzhang@gxu.edu.cn, xzhang@amss.ac.cn}
	
	\date{}
	
	\begin{abstract}
		Geodesic equations are solved when at least two of $\theta$, $\phi$ and $\psi$ are constant, or $r$ is constant, on scalar flat metrics of Eguchi-Hanson type. They can also be solved also on Eguchi-Hanson metrics which are Ricci flat if only $\phi$ is constant. However, the explicit solution of the geodesic equations is not available yet if only $\psi$ is constant.
	\end{abstract}

	\maketitle \pagenumbering{arabic}
	
	\mysection{Introduction}\ls
	
Geodesics are the fundamental geometric object in differential geometry. They also reveal nature of gravitational phenomena in curved space-time. In general relativity, null geodesics are interpreted as trajectories of light, time-like geodesics are interpreted as trajectories of massive particles, and they are studied extensively in many exact spacetimes, cf. \cite{C, CM} and references therein. Besides enormous study of geodesics in Riemannian manifolds, Battista and Esposito recently investigated geodesic motion on Euclidean Schwarzschild metrics over $R^2 \times S^2$ and obtained its explicit form in terms of incomplete elliptic integrals of the first, the second and the third kinds \cite{BE}. As Euclidean Schwarzschild metrics belong to gravitational instantons, which play important roles in Euclidean approach of quantum gravity \cite{GH}, it is interesting to investigate geodesic motion on other gravitational instantons such as Eguchi-Hanson metrics \cite{EH1, EH2}. This is the goal of our paper.

Eguchi-Hanson metrics are complete 4-dimensional Ricci flat, anti-self-dual ALE Riemannian metrics on $R_{\geq 0} \times P^3$. The more general metrics of Eguchi-Hanson type were constructed by LeBurn \cite{L} using the method of algebraic geometry \cite{L} and by the second author solving an ordinary differential equation \cite{Z}. They are scalar flat on $R_{\geq 0} \times S^3 /Z_d$ $(d >2)$ and provide counter-examples of Hawking and Pope's generalized positive action conjecture \cite{HP}. It is important that gravitational instantons always have removable singularity. And it occurs at the largest positive root of potential functions for metrics of Eguchi-Hanson type. The geodesic completeness has not been proved in the literature for these types of metrics, as pointed out in \cite{BE}. This motivates us to solve geodesic equations passing through the removable singularity.

The paper is organized as follows. In Section \ref{2}, we provide a brief introduction to Eguchi-Hanson metrics and scalar flat metrics of Eguchi-Hanson type. We also give geodesic equations for metrics of Eguchi-Hanson with general radial potential functions. In Section \ref{3}, we solve geodesic equations on scalar flat metrics of Eguchi-Hanson type in the following cases: (1) Subsection \ref{subsec1}, where $\theta$, $\phi$, $ \psi$ are constant, geodesics pass through the removable singularity; (2) Subsection \ref{subsec2}, where $\phi$, $\psi$ are constant, geodesics pass through the removable singularity; (3) Subsection \ref{subsec3}, where $\theta$, $\phi$ are constant, geodesics do not pass through the removable singularity. If geodesics pass through the removable singularity, $\psi$ must be constant and the geodesic equations reduce to Subsection \ref{subsec1}; (4) Subsection \ref{subsec4}, where $\theta\in (0, \pi)$, $\psi$ are constant. It yields that $\theta=\frac{\pi}{2}$ if $\phi$ is not constant; (5) Subsection \ref{subsec5}, where $r$ is constant, geodesics pass through the removable singularity; (6) Subsection \ref{subsec6}, where $\phi$ is constant. It yields that either $\psi$ or $\theta$ is constant, and the geodesic equations reduce to either Subsection \ref{subsec2} or Subsection \ref{subsec3}. In Section \ref{4}, we solve geodesic equations on Eguchi-Hanson metrics, where only $\theta\in (0, \pi)$ is constant and geodesics do not pass through the removable singularity. If geodesics pass through the removable singularity, then $\theta=\frac{\pi}{2}$, $\psi$ is constant and the geodesic equations reduce to Subsection \ref{subsec4}. In Appendix \ref{5}, we provide an brief introduction to incomplete elliptic integrals of the first, the second and the third kinds.
	
We point out that the explicit solution of geodesic equations is not available yet if only $\psi$ is constant.

	\mysection{Metrics of Eguchi-Hanson type and geodesic equations} \label{2}
	\ls
	
	In this section, we introduce the metrics of Eguchi-Hanson type and provide their geodesic equations. Let $\sigma_{1}$, $\sigma_{2}$, and $\sigma_{3}$ be the Cartan-Maurer one-forms for $SU(2)\cong S^3$, defined by
	\begin{align*}
		\sigma_{1}&=\frac{1}{2}\big(\sin\psi d\theta-\sin\theta \cos\psi d\phi \big),\\
		\sigma_{2}&=\frac{1}{2}\big(-\cos\psi d\theta-\sin\theta \sin\psi d\phi \big),\\
		\sigma_{3}&=\frac{1}{2}\big(d \psi+\cos\theta d\phi \big).
	\end{align*}
They satisfy
	\begin{align*}
		d\sigma_{1}=2\sigma_{2}\wedge\sigma_{3},\quad d\sigma_{2}=2\sigma_{3}\wedge  \sigma_{1},\quad d\sigma_{3}=2\sigma_{1}\wedge  \sigma_{2}.
	\end{align*}
	Metrics of Eguchi-Hanson type are given by
	\beq\label{EH-0}
	ds^2 = f^{-1} dr ^2 +r ^2 \Big( \sigma _1 ^2 +\sigma _2 ^2 +f \sigma _3 ^2 \Big),
	\eeq
	where $f\geq0$ is a smooth function, which is referred as the {\it potential} function.

If
\beq\label{EH-}
	f=1-\frac{B}{r ^4}
\eeq
for some $B>0$, and
	\beq
r \geq \sqrt[4]{B},\quad 0 \leqslant \theta \leqslant \pi, \quad 0 \leqslant \phi \leqslant 2\pi, \quad 0 \leq \psi \leq 2\pi,
	\eeq
the metrics (\ref{EH-0}) are Eguchi-Hanson metrics \cite{EH1, EH2}, which are Ricci flat, geodesically complete and asymptotically local Euclidean, and $\sqrt[4]{B}$ is the removable singularity. The underground manifold is, topologically,
	\begin{align*}
		R_{\geq 0} \times SU(2)/Z_2 \cong R_{\geq 0} \times SO(3) \cong R_{\geq 0} \times P _3.
	\end{align*}
	
Let $n \geq 2$ be a natural number. If
	\beq\label{f0}
	f=1-\frac{2A}{r ^2}-\frac{B}{r ^4},\quad A=-\frac{n-2}{2}\sqrt{\frac{B}{n-1}},
	\eeq
for some $B>0$, and
	\begin{align}
		r\geq \sqrt[4]{\frac{B}{n-1}},  \quad 0 \leqslant \theta \leqslant \pi, \quad 0 \leqslant \phi \leqslant 2\pi, \quad 0 \leq \psi \leq \frac{4\pi}{n},
	\end{align}
the metrics (\ref{EH-0}) are geodesically complete, asymptotically local Euclidean, and $\sqrt[4]{\frac{B}{n-1}}$ is the removable singularity. The underground manifold is, topologically,
\begin{align*}
R_{\geq 0} \times SU(2)/Z_n \cong R_{\geq 0} \times S^3/Z_n.
\end{align*}
With $A$ given by (\ref{f0}), we have
\begin{align*}
r^4 -2A r^2 -B=\left(r^2 - \sqrt{\frac{B}{n-1}}  \right)\left(r^2 +\sqrt{(n-1)B  }   \right).
\end{align*}
Thus $\sqrt[4]{\frac{B}{n-1}}$ is the largest positive simple root of $f$. This fact is crucial to solve radial geodesics passing through the removable singularity.

The nonvanishing metric components of (\ref{EH-0}) are
	\begin{align*}
		g_{rr}&=f^{-1} , &g^{rr}&=f,\\
		g_{\theta \theta}&=\frac{r^2}{4} ,&  g^{\theta \theta}&=\frac{4}{r ^2},\\
		g_{\phi\phi}&=\frac{r^2}{4}\left(f\cos^2\theta+\sin^2\theta\right) , &g^{\phi\phi}&=\frac{4}{r ^2\sin^2\theta},\\
		g_{\phi\psi}&=g_{\psi\phi}=\frac{r^2}{4}f\cos\theta,&g^{\phi\psi}&=g^{\psi\phi}=-\frac{4\cos\theta}{r ^2\sin^2\theta},\\
		g_{\psi\psi}&=\frac{r^2}{4}f,
		&g^{\psi\psi}&=\frac{4}{r ^2}\left(\cot^2\theta+\frac{1}{f}\right).
	\end{align*}
The nonvanishing Christoffel symbols are
	\begin{align*}
		\Gamma^r_{rr}&=-\frac{f'}{2f},
                     & \Gamma^r_{\theta\theta}&=-\frac{r}{4}f,\\
		\Gamma^r_{\phi\phi}&=-\frac{f}{4}\left(\frac{\left(r^2f\right)'}{2}\cos^2\theta+r\sin^2\theta\right),
                           & \Gamma^r_{\phi\psi}&=-\frac{\left(r^2f\right)'}{8f^{-1}}\cos\theta,\\
		\Gamma^r_{\psi\psi}&=-\frac{f \left(r^2f\right)'}{8},
                           & \Gamma^\theta_{r\theta}&=\Gamma^\phi _{r\phi}=\frac{1}{r},\\
		\Gamma^\theta_{\phi \phi}&=\left(f-1\right)\sin\theta \cos\theta ,
                     & \Gamma^\theta _{\phi \psi}&=\frac{f}{2}\sin\theta,\\
		\Gamma^\phi_{\theta \phi}&=\left(1-\frac{f}{2}\right)\cot\theta,
                     & \Gamma^\phi_{\theta \psi}&=-\frac{f}{2\sin\theta},\\
		\Gamma^\psi_{r \phi}&=\frac{f'}{2f}\cos\theta,
                     & \Gamma^\psi_{r\psi}&=\frac{1}{r}+\frac{f'}{2f},\\
		\Gamma^\psi_{\theta\phi}&=\frac{\left(f-1\right)\cos^2\theta-1}{2\sin\theta},
                     & \Gamma^\psi_{\theta \psi}&=\frac{f\cot\theta}{2},
	\end{align*}
Thus geodesic equations of (\ref{EH-0}) for parameter $t$ are
	\begin{align}
\frac{d^2r}{dt^2}&-\frac{f'}{2f}\Big(\frac{dr}{dt}\Big)^{2}-\frac{rf}{4}\Big(\frac{d\theta}{dt}\Big)^{2}
		-\frac{f}{4}\Big(r\sin^2\theta+\frac{(r^2f )' }{2}\cos^2\theta\Big)\Big(\frac{d\phi}{dt}\Big)^{2}	\notag\\
		&-\frac{f(r^2f) '}{4}\cos\theta\frac{d\phi}{dt}\frac{d\psi}{dt}-\frac{f(r^2f) '}{8}\Big(\frac{d\psi}{dt}\Big)^{2}=0,\label{eqa}\\
\frac{d^2\theta}{dt^2}&+\frac{2}{r}\frac{dr}{dt}\frac{d\theta}{dt}
        +(f-1)\sin\theta \cos\theta\Big(\frac{d\phi}{dt}\Big)^{2}+f\sin\theta\frac{d\phi}{dt}\frac{d\psi}{dt}=0,\label{eqb}\\
\frac{d^2\phi}{dt^2}&+\frac{2}{r}\frac{dr}{dt}\frac{d\phi}{dt}-(f-2)\cot\theta\frac{d\phi}{dt}\frac{d\theta}{dt}
		-\frac{f}{\sin\theta}\frac{d\theta}{dt}\frac{d\psi}{dt}=0,	 \label{eqc}\\
\frac{d^2\psi}{dt^2}&+\frac{f'}{f}\cos\theta\frac{dr}{dt}\frac{d\phi}{dt}+\frac{(r^2f)'}{r^2f}\frac{dr}{dt}\frac{d\psi}{dt}
		+\frac{(f-1)\cos^2\theta-1}{\sin\theta}\frac{d\theta}{dt}\frac{d\phi}{dt}	\notag\\
		&+f\cot\theta\frac{d\theta}{dt}\frac{d\psi}{dt}=0.\label{eqd}
	\end{align}	

Throughout the paper, we denote $F(\alpha, k^2)$, $E(\alpha, k^2)$, and $\Pi(h,\alpha, k^2)$ the incomplete elliptic integrals of the first, second, and third kind, where $ 0<k<1 $, $ h\in\mathbb{C} $ (cf. Appendix), and
\begin{align*}
\varepsilon=\pm1, \quad r_0= \sqrt[4]{\frac{B}{n-1}}
\end{align*}
for $B>0$, $n\geq 2.$
	
	\mysection{Geodesics on scalar flat metrics of Eguchi-Hanson type}\label{3}
	\ls
	
	In this section, we shall solve the geodesic equations for the following cases when $f$ is given by (\ref{f0}).
	
	\subsection{Geodesics for constant $\theta$, $\phi$, $ \psi$}\label{subsec1}
	\ls
	
	The geodesic equations reduce to
	\begin{equation}\label{k11}
		\frac{d^2r}{dt^2}-\frac{f'}{2f}\left(\frac{dr}{dt}\right)^{2}=0.
	\end{equation}
	
	\begin{thm}\label{thm3-1}
		The geodesics for scalar flat metrics of Eguchi-Hanson type with constant
		$\theta$, $\phi$ and $ \psi$ and passing through $r_0$ with conditions
		\begin{align*}
			\lim\limits_{r\rightarrow r_0}  t =t_1, \quad \lim\limits_{r\rightarrow r_0} \frac{1}{\sqrt{f}}\frac{dr}{dt} = r_1
		\end{align*}
		satisfy
		\begin{align*}
			t(r)=& t_1+\frac{\varepsilon}{r_1} \Big[\frac{\sqrt{r^4+(n-2)r_0^2r^2-B }}{r} \\
			     & -\sqrt{n}r_0 E\left(\alpha,k^2\right)+\frac{r_0}{\sqrt{n}}F\left(\alpha, k^2\right)\Big],
		\end{align*}
		where
\begin{align*}
\alpha=\arcsin\sqrt{1-\frac{r_0^2}{r^2}}, \quad k^2=1-\frac{1}{n}.
\end{align*}
	\end{thm}
	\pf The geodesic equation (\ref{k11}) implies that
	\begin{align*}
		\frac{d}{dt}\Big(\frac{1}{\sqrt{f}}\Big|\frac{dr}{dt}\Big|\Big)=0.
	\end{align*}
	Thus
	\begin{align*}
		\frac{dt}{dr}=\frac{  \varepsilon}{r_1\sqrt{f}}.
	\end{align*}
	Changing variable  $ u=r^2 $, we obtain
	\begin{align*}
		\frac{dt}{du} =\frac{ \varepsilon}{2r_1}\frac{ u}{\sqrt{u\big( u-r_0^2\big) \big( u+\sqrt{(n-1)B}\big) }}.
	\end{align*}
	The theorem follows by integrating it from $ r_0^2$  to $ r^2 $.\qed
	
	\subsection{Geodesics for constant $\phi$, $ \psi$}\label{subsec2}
	\ls
	
	The geodesic equations reduce to
	\begin{align}
		&\frac{d^2r}{dt^2} -\frac{f'}{2f}\Big(\frac{dr}{dt}\Big)^{2}-\frac{rf}{4}\Big(\frac{d\theta}{dt}\Big)^{2}=0,\label{eq001}\\
		&\frac{d^2\theta}{dt^2}+\frac{2}{r}\frac{dr}{dt}\frac{d\theta}{dt}=0.\label{eq002}
	\end{align}
	
	\begin{thm}\label{thm3-2}
Let constant $r_1$, $\theta _0$ satisfy
\begin{align*}
 r_1>\frac{r_0\theta_0 }{2}.
\end{align*}
The geodesics for scalar flat metrics of Eguchi-Hanson type with constant $\phi$, $ \psi$ and passing through $r_0$ with conditions
		\begin{align*}
			\lim\limits_{r\rightarrow r_0}  t =t_1, \quad
			\lim\limits_{r\rightarrow r_0} \theta=\theta_1, \quad
			\lim\limits_{r\rightarrow r_0}\frac{d\theta}{dt}=\theta_0, \quad
			\lim\limits_{r\rightarrow r_0} \frac{1}{\sqrt{f}} \left| \frac{dr}{dt} \right| =\sqrt{r_1^2-\frac{r_0^2\theta_0^2}{4}}
		\end{align*}
		satisfy
		\begin{align*}
				t(r) =&t_1+\frac{\varepsilon}{r_1} \Big[\frac{2r_1\sqrt{r^4+(n-2)r_0^2r^2-B }}{\sqrt{4r_1^2r^2-r_0^4\theta_0^2}}\\
			          &-\sqrt{n}r_0 E\left(\alpha,k^2\right)+\frac{r_0}{\sqrt{n}}F\left(\alpha, k^2\right)\Big],\\
           \theta(r) =&\theta_1 +\frac{\varepsilon r_0\,\theta_0 }{ \sqrt{n}r_1}F\left(\alpha,k^2\right),
		\end{align*}
		where
\begin{align*}
\alpha=\arcsin\sqrt \frac{4r_1^2	r^2-4r_0^2r_1^2}{4r_1^2	r^2-r_0^4\theta_0^2 },\quad
k^2= 1-\frac{1}{n}+\frac{r_0^2\theta_0^2}{4nr_1^2}.
\end{align*}
	\end{thm}
	\pf The geodesic equation (\ref{eq002}) implies that
	\begin{align*}
		\frac{d}{dt}\ln \Big( r^2 \Big| \frac{d\theta}{dt}\Big| \Big)=0.
	\end{align*}
	Thus
	\begin{align*}
		\frac{d\theta}{dt}=\frac{\varepsilon r_0^2 \theta_0}{r^{2}}.
	\end{align*}
	Substituting it into (\ref{eq001}), we obtain
	\begin{align*}
		\frac{d}{dt}\Big(\frac{1}{f} \Big(\frac{dr}{dt}\Big)^{2}+\frac{r_0^4\theta_0^{2}}{4r^{2}}\Big)=0.
	\end{align*}
	Thus
	\begin{align*}
		\frac{1}{f} \Big(\frac{dr}{dt}\Big)^{2}+\frac{r_0^4\theta_0^{2}}{4r^{2}}=r_1^2.
	\end{align*}
	Therefore
	\begin{align*}
		\frac{d\theta}{dr} &=\frac{d\theta}{dt} \Big/ \frac{dr}{dt}=\frac{\varepsilon r_0^2\theta_0}{r_1r^2\sqrt{\Big(1-\frac{r_0^4\theta_0^2}{4r_1^2r^2}\Big)f}},\\
			\frac{dt}{dr} &=\frac{\varepsilon }{r_1\sqrt{\Big(1-\frac{r_0^4\theta_0^2}{4r_1^2r^2}\Big)f}}.
	\end{align*}
	Changing variable $ u=r^2 $, we obtain
	\begin{align*}
		\frac{d\theta}{du} &=\frac{\varepsilon r_0^2\theta_0}{2r_1\sqrt{P(u)}},\\
			\frac{dt}{du} &=\frac{\varepsilon u}{2r_1\sqrt{P(u)}},
	\end{align*}
	with
	\begin{equation*}
		P(u)=\left( u-\frac{r_0^4\theta_0^2}{4r_1^2}\right)\left( u-r_0^2\right) \left( u+\sqrt{(n-1)B} \right) .
	\end{equation*}
	The theorem follows by integrating them from $ r_0^2$  to $ r^2 $.\qed

	\subsection{Geodesics for constant $\theta$, $\phi$}\label{subsec3}
	\ls
	
	The geodesic equations reduce to
	\begin{align}
		&\frac{d^2r}{dt^2} -\frac{f'}{2f}\left(\frac{dr}{dt}\right)^{2}-	 \frac{f\left(r^2f\right)'}{8}\left(\frac{d\psi}{dt}\right)^{2}=0,\label{eq01}\\
		&\frac{d^2\psi}{dt^2}+\frac{\left(r^{2}f\right)'}{r^{2}f}\frac{dr}{dt}\frac{d\psi}{dt} =0.\label{eq02}
	\end{align}

Let $r_1$, $\psi_0$ be constant and $ r_1\neq 0$. Denote
\begin{equation*}
r_{+} =\frac{1}{2\sqrt{2}} \sqrt{4(2-n)r_0^2+\frac{\psi_0^2}{r_1^2}+\sqrt{\left( 4(2-n)r_0^2+\frac{\psi_0^2}{r_1^2}\right) ^2+64B}}.
\end{equation*}

\begin{lem}\label{lemma_r+}
With the above notations, $ r_+\geq r_0$, and $ r_+=r_0 $  if and only if $ \psi_0=0 $.
\end{lem}
\pf A straightforward computation. \qed

\begin{thm}\label{thm3-3-1}
Let $ r \geq r_+> r_0$. The geodesics for scalar flat metrics of Eguchi-Hanson type with constant $ \theta$, $\phi$ and conditions
\begin{align*}
	\lim\limits_{r\rightarrow r_+}  t =t_1, \quad
\lim\limits_{r\rightarrow r_+}  \psi=\psi_1,
		\quad 	\lim\limits_{r\rightarrow  r_+} & r^2 f  \frac{d\psi}{dt}=\psi_0, \\
	\lim\limits_{r\rightarrow r_+}	\left( \frac{1}{f}\left(\frac{dr}{dt}\right)^{2}+\frac{\psi_0^{2}}{4r^{2}f}\right) & =r_1^2
\end{align*}
satisfy
			\begin{align*}
t(r)=&t_1+ \frac{\sqrt{r^4-\left( \frac{\psi_0^2}{4r_1^2}+(2-n)r_0^2\right) r^2-B }}{\varepsilon r_1 r}\\
			&-\frac{\sqrt[4]{\left( 4(2-n)r_0^2r_1^2+\psi_0^2\right) ^2+64Br_1^4}}{2\varepsilon r_1^2} E\left(\alpha,k^2\right)\\
			&+\frac{2\varepsilon r_+^2}{\sqrt[4]{\left( 4(2-n)r_0^2r_1^2+\psi_0^2\right) ^2+64Br_1^4}}F\left(\alpha, k^2\right),\\
\psi(r)=&\psi_1+\frac{2\varepsilon\psi_0\,r_{+}^2\,\Pi(h_1,\alpha,k^2)}{n(r_{+}^2-r_0^2)\sqrt[4]{\left( 4(2-n)r_0^2r_1^2+\psi_0^2\right) ^2+64Br_1^4}} \notag\\
		&	+\frac{2\varepsilon(n-1)\psi_0\,r_{+}^2\,\Pi(h_2,\alpha,k^2)}{n(r_{+}^2+\sqrt{(n-1)B})\sqrt[4]{\left( 4(2-n)r_0^2r_1^2+\psi_0^2\right) ^2+64Br_1^4}} ,
		\end{align*}
		where
		\begin{align*}
		\alpha &=\arcsin\sqrt {1-\frac{4(2-n)r_0^2r_1^2+\psi_0^2+\sqrt{\left( 4(2-n)r_0^2r_1^2+\psi_0^2\right) ^2+64Br_1^4}}{8r_1^2r^2}},\\
		k^2 &=\frac{1}{2}\left(1- \frac{4(2-n)r_0^2r_1^2+\psi_0^2}{\sqrt{\left( 4(2-n)r_0^2r_1^2+\psi_0^2\right) ^2+64Br_1^4}}\right) ,\\
		h_1 &=\frac{-8r_0^2r_1^2}{\psi_0^2-4nr_0^2r_1^2+\sqrt{\left( 4(2-n)r_0^2r_1^2+\psi_0^2\right) ^2+64Br_1^4}} ,\\
		h_2 &=\frac{8\sqrt{(n-1)B}r_1^2}{\psi_0^2+4nr_0^2r_1^2+\sqrt{\left( 4(2-n)r_0^2r_1^2+\psi_0^2\right) ^2+64Br_1^4}}.
		\end{align*}
\end{thm}
\pf
	The geodesic equation (\ref{eq02}) implies that
	\begin{align*}
		\frac{d}{dt}\ln\left(r^2f\left| \frac{d\psi}{dt}\right| \right)=0.
	\end{align*}
	Thus
	\begin{align}\label{a1}
		\frac{d\psi}{dt}=\frac{\varepsilon\psi_0}{r^{2}f}.
	\end{align}
Substituting it into (\ref{eq01}), we obtain
	\begin{align*}
		\frac{d}{dt}\left(\frac{1}{f}\left(\frac{dr}{dt}\right)^{2}+\frac{\psi_0^{2}}{4r^{2}f}\right)=0.
	\end{align*}
	Thus
	\begin{align*}
		\frac{1}{f}\left(\frac{dr}{dt}\right)^{2}+\frac{\psi_0^{2}}{4r^{2}f}=r_1^2.
	\end{align*}
Therefore
\begin{align*}
	\frac{d\psi}{dr}& =\frac{d\psi}{dt} \Big/ \frac{dr}{dt}=\frac{\varepsilon\psi_0}{r_1r^2f\sqrt{f-\frac{\psi_0^2}{4r_1^2r^2}}},\\
	\frac{dt}{dr} &=\frac{\varepsilon}{r_1\sqrt{f-\frac{\psi_0^2}{4r_1^2r^2}}}.
\end{align*}
Changing variable $ u=r^2 $, we obtain
\begin{align}
	\frac{d\psi}{du}& =\frac{\varepsilon\psi_0\, u^2}{2r_1\left( u-r_0^2\right) \left( u+\sqrt{(n-1)B}\right) \sqrt{P(u)}}, \label{n 91}\\
	\frac{dt}{du} &=\frac{\varepsilon u}{2r_1 \sqrt{P(u)}}, \label{n 92}
\end{align}
with
\begin{align*}
	P(u) =&u^3-\left( \frac{\psi_0^2}{4r_1^2}+(2-n)r_0^2\right) u^2-Bu \\
	=&u\left( u-u_0\right) \left( u-u_1\right) ,
\end{align*}
where
\begin{align*}
	&u_0=\frac{4(2-n)r_0^2r_1^2+\psi_0^2+\sqrt{\left( 4(2-n)r_0^2r_1^2+\psi_0^2\right) ^2+64Br_1^4}}{8r_1^2},\\
	&u_1=\frac{4(2-n)r_0^2r_1^2+\psi_0^2-\sqrt{\left( 4(2-n)r_0^2r_1^2+\psi_0^2\right) ^2+64Br_1^4}}{8r_1^2}.
\end{align*}
The theorem follows by integrating (\ref{n 91}) and  (\ref{n 92}) from $ u_0$  to $ r^2 $. \qed

\begin{thm}\label{thm3-3-2}
The geodesics for scalar flat metrics of Eguchi-Hanson type with constant $ \theta$, $\phi$ and passing through $r_0$ with conditions
\begin{align*}
\lim\limits_{r\rightarrow r_0}  t =t_1, \quad
\lim\limits_{r\rightarrow r_0} \psi =\psi_1, \quad
\lim\limits_{r\rightarrow r_0} \frac{1}{\sqrt{f}}\frac{dr}{dt} = r_1
\end{align*}
satisfy
\begin{align*}
t(r) =& t_1+\frac{\varepsilon}{r_1} \Big[\frac{\sqrt{r^4+(n-2)r_0^2r^2-B }}{r}\\
      & -\sqrt{n}r_0 E\left(\alpha,k^2\right)+\frac{r_0}{\sqrt{n}}F\left(\alpha, k^2\right)\Big],\\
\psi(r) =&\psi_1,
\end{align*}
where
\begin{align*}
\alpha=\arcsin\sqrt{1-\frac{r_0^2}{r^2}}, \quad k^2=1-\frac{1}{n}.
\end{align*}
\end{thm}
\pf Lemma \ref{lemma_r+} implies that $ \psi_0=0 $. Then (\ref{a1}) gives that $ \frac{d \psi}{dt} =0$, and it reduces to Theorem \ref{thm3-1}. \qed
	
	\subsection{Geodesics for constant $ \theta \in (0, \pi)$,  $\psi$ }\label{subsec4}
	\ls
	
The geodesic equations (\ref{eqb}), (\ref{eqd}) give that
	\begin{align*}
		\left(f-1\right)\sin\theta \cos\theta \left(\frac{d\phi}{dt}\right)^{2}=0,\quad
		\frac{f'}{f}\cos\theta\frac{dr}{dt}\frac{d\phi}{dt}=0.
	\end{align*}
They imply either $ \dfrac{d\phi}{dt}=0 $ or $ \theta=\dfrac{\pi}{2} $. It reduces to Theorem \ref{thm3-1} if $\dfrac{d\phi}{dt}=0$.

Now we focus on the case $ \theta=\dfrac{\pi}{2} $. The geodesic equations reduce to
	\begin{align}
		\frac{d^2r}{dt^2}&-\frac{f'}{2f}\left(\frac{dr}{dt}\right)^{2}-\frac{rf}{4}\left(\frac{d\phi}{dt}\right)^{2}=0,\label{eq3-4-11}\\
		\frac{d^2\phi}{dt^2}&+\frac{2}{r}\frac{dr}{dt}\frac{d\phi}{dt}=0,\label{eq3-4-2}
	\end{align}

\begin{thm}\label{thm3-4}
Let constant $r_1$, $\phi_0$ satisfy
\begin{align*}
r_1>\frac{r_0\phi_0 }{2}.
\end{align*}
The geodesics for scalar flat metrics of Eguchi-Hanson type with constant $\theta \in (0,\pi)$, $ \psi$, nonconstant $\phi$ and passing through $r_0$ with conditions
		\begin{align*}
			\lim\limits_{r\rightarrow r_0}  t =t_1, \quad \lim\limits_{r\rightarrow r_0} \phi=\phi_1, \quad
			\lim\limits_{r\rightarrow r_0}\frac{d\phi}{dt}=\phi_0, \quad
			\lim\limits_{r\rightarrow r_0} \frac{1}{\sqrt{f}} \left| \frac{dr}{dt} \right| =\sqrt{r_1^2-\frac{r_0^2\phi_0^2}{4}}
		\end{align*}
		satisfy
		\begin{align*}
		t(r)=&t_1+\frac{\varepsilon}{r_1} \Big[\frac{2r_1\sqrt{r^4+(n-2)r_0^2r^2-B }}{\sqrt{4r_1^2r^2-r_0^4\theta_0^2}}\\
		     &-\sqrt{n}r_0 E\left(\alpha,k^2\right)+\frac{r_0}{\sqrt{n}}F\left(\alpha, k^2\right)\Big],\\
     \theta(r) =&\frac{\pi}{2}, \\
     \phi(r)=&\phi_1 +\frac{\varepsilon r_0\,\phi_0 }{ \sqrt{n}r_1}F\left(\alpha,k^2\right),
		\end{align*}
		where
\begin{align*}
\alpha=\arcsin\sqrt\frac{4r_1^2	r^2-4r_0^2r_1^2}{4r_1^2	r^2-r_0^4\phi _0^2 },\quad
k^2= 1-\frac{1}{n}+\frac{r_0^2\phi_0^2}{4nr_1^2}
\end{align*}
	\end{thm}
\pf Same as the proof of Theorem \ref{thm3-2}.\qed
	
	\subsection{Geodesics for constant  $ r $}\label{subsec5}
	\ls
	
	The geodesic equations reduce to
	\begin{align}
&\frac{d^2\theta}{dt^2}+\left(f-1\right)\sin\theta \cos\theta\left(\frac{d\phi}{dt}\right)^{2}
                       +f\sin\theta\frac{d\phi}{dt}\frac{d\psi}{dt}=0	 ,\label{Eq41}\\
&\frac{d^2\phi}{dt^2}-\left(f-2\right)\cot\theta\frac{d\phi}{dt}\frac{d\theta}{dt}
                     -\frac{f}{\sin\theta}\frac{d\theta}{dt}\frac{d\psi}{dt}=0,\label{Eq43}\\
&\frac{d^2\psi}{dt^2}+\frac{\left(f-1\right)\cos^2\theta-1}{\sin\theta}\frac{d\theta}{dt}\frac{d\phi}{dt}
                     +f\cot\theta\frac{d\theta}{dt}\frac{d\psi}{dt}=0,\label{Eq42}\\
&\frac{r}{4}f\left(\frac{d\theta}{dt}\right)^{2}+\frac{r}{4}f\sin^2\theta\left(\frac{d\phi}{dt}\right)^{2}=0,\label{Eq44}
	\end{align}

As $r=r_0$ is the coordinate origin, we assume $r>r_0$.

	\begin{thm}\label{thm4}
			The geodesics for scalar flat metrics of Eguchi-Hanson type with constant $ r>r_0 $ and conditions
\begin{align*}
\lim\limits_{r\rightarrow r_0} \theta =\theta_1, \quad \lim\limits_{r\rightarrow r_0} \phi =\phi_1, \quad
\lim\limits_{r\rightarrow r_0} \psi =\psi_1, \quad \lim\limits_{r\rightarrow r_0} \frac{d\psi}{dt} =\psi_0
\end{align*}
satisfy
\begin{align*}
\theta(t)=\theta_1, \quad \phi(t)=\phi_1, \quad \psi(t)=\psi_1 +\psi_0 t.
\end{align*}
	\end{thm}
	\pf The geodesic equation (\ref{Eq44}) implies that $ \dfrac{d\theta}{dt}=\dfrac{d\phi}{dt}=0 $. Then (\ref{Eq42}) gives
$ \dfrac{d^2\psi}{dt^2}=0 $ and the theorem follows. \qed

	\subsection{Geodesics for constant  $ \phi$}\label{subsec6}
	\ls
	
The geodesic equation (\ref{eqc}) gives
	\begin{equation*}
		\frac{f}{\sin\theta}\frac{d\theta}{dt}\frac{d\psi}{dt}=0.
	\end{equation*}
Therefore $ \dfrac{d\psi}{dt}=0 $ or $ \dfrac{d\theta}{dt}=0 $, and it reduces to the theorems in Subsection \ref{subsec2} or Subsection \ref{subsec3} respectively.

	\mysection{Geodesics for constant $\theta$ on Eguchi-Hanson metrics}\label{4}
	
	In this section, we solve the geodesic equations for constant $\theta\in\left( 0, \pi\right) $ on Eguchi-Hanson metrics with $f$ given by (\ref{EH-}). The geodesic equations reduce to
	\begin{align}
		\frac{d^2r}{dt^2}&-\frac{f'}{2f}\left(\frac{dr}{dt}\right)^{2}-\frac{f}{4}\left(r\sin^2\theta+\frac{\left(r^2f \right)' }{2}\cos^2\theta\right)\left(\frac{d\phi}{dt}\right)^{2}
		\notag
		\\
		&-\frac{f\left(r^2f \right) '}{4}\cos\theta\frac{d\phi}{dt}\frac{d\psi}{dt}
		-\frac{f\left(r^2f \right) '}{8}\left(\frac{d\psi}{dt}\right)^{2}=0,\label{eq21}\\
		f\frac{d\phi}{dt}&\frac{d\psi}{dt}+(f-1) \cos\theta\left(\frac{d\phi}{dt}\right)^{2}=0,\label{eq22}\\
		\frac{d^2\phi}{dt^2}&+\frac{2}{r}\frac{dr}{dt}\frac{d\phi}{dt}=0,\label{eq23}	\\
		\frac{d^2\psi}{dt^2}&+\frac{f'}{f}\cos\theta\frac{dr}{dt}\frac{d\phi}{dt}+\frac{\left( r^2f\right)'}{r^2f}\frac{dr}{dt}\frac{d\psi}{dt}=0. \label{eq24}
	\end{align}
Equation (\ref{eq22}) implies that either $ \dfrac{d\phi}{dt}=0 $ or
	\begin{equation}\label{eqf1}
		\left(f-1\right) \cos\theta\frac{d\phi}{dt}+f\frac{d\psi}{dt}=0.
	\end{equation}
It reduces to the theorems in Subsection \ref{subsec3} if $ \dfrac{d\phi}{dt}=0 $.

Now we focus on the case $ \dfrac{d\phi}{dt}\neq 0 $. Then (\ref{eqf1}) holds.
\begin{lem}\label{T0-1}
Let function $T(x)$
\begin{align}
T(x)=\frac{x \left(27\sin^2\theta- 2x^2 -9\right)}{2(x^2+3)^\frac{3}{2}}, \quad x \geq 1
\end{align}
for fixed $\theta \in (0, \pi)$, then it is monotonically decaying and
\begin{align}
-1<T(x)\leq 1, \quad x=1 \Llrw \theta =\frac{\pi}{2}.  \label{T0-2}
\end{align}
\end{lem}
\pf It is straightforward that
\begin{align*}
T'(x)=-\frac{27\left[ (2x^2-3)\sin ^2 \theta+1\right] }{2 (x^2+3)^\frac{5}{2}}\leq 0,
\end{align*}
and
\begin{align*}
T(1)=\frac{27 \sin^2 \theta -11}{16} \leq 1,\quad T(\infty)=-1.
\end{align*}
Therefore the lemma follows. \qed

Let $r_1$, $ \phi_0 $ be constant such that
\begin{align}
r_1\neq 0, \quad \left| \phi_0\right| \geq\frac{2|r_1|}{\sqrt[4]{B}}. \label{T0-3}
\end{align}
By Lemma \ref{T0-1}, we can define
\begin{align*}
T=T\left(\frac{\sqrt{B}\phi_0 ^2}{4r_1 ^2} \right)=\frac{\sqrt{B}\phi_0^2\left(216r_1^4\sin^2\theta-B\phi_0^4-72r_1^4\right)}{\sqrt{(B\phi_0^4+48r_1^4)^3}}.
\end{align*}
and
\begin{align*}
\bar{r}_{0}=\frac{1}{2\sqrt{3}} \sqrt{B\frac{\phi_0^2}{r_1^2}-2\sqrt{B^2\frac{\phi_0^4}{r_1^4}+48B}\cos\left( \frac{\eta}{3}+ \frac{2\pi}{3}\right) }, \quad \eta=\arccos T.
\end{align*}

\begin{lem}\label{lemm4-1}
	Suppose that (\ref{T0-3}) holds, then
\begin{align*}
	\bar{r} _0\geq\sqrt[4]{B}.
\end{align*}
Equality holds if and only if
\begin{align*}
\left| \phi_0\right| =\frac{2|r_1|}{\sqrt[4]{B}},\quad \theta=\frac{\pi}{2}.
\end{align*}
\end{lem}
\pf  It is obvious that  $ \eta\in[0, \pi) $.
A straightforward calculation shows that
\begin{align*}
-1< \cos\left( \frac{\eta}{3}+ \frac{2\pi}{3}\right)\leq-\frac{1}{2}.
\end{align*}
Thus
 \begin{align*}
 	\bar{r}_{0}&=\frac{1}{2\sqrt{3}} \sqrt{B\frac{\phi_0^2}{r_1^2}-2\sqrt{B^2\frac{\phi_0^4}{r_1^4}+48B}\cos\left( \frac{\eta}{3}+ \frac{2\pi}{3}\right) }\\
 	&\geq\frac{1}{2\sqrt{3}} \sqrt{B\frac{\phi_0^2}{r_1^2}+\sqrt{B^2\frac{\phi_0^4}{r_1^4}+48B}}\\
 	&\geq \sqrt[4]{B}.
 \end{align*}
Equality holds if and only if
\begin{align*}
\frac{\phi_0^2}{r_1^2}=\frac{4}{\sqrt{B}}, \quad \eta=0 \Llrw \left| \phi_0\right| =\frac{2|r_1|}{\sqrt[4]{B}},\quad \theta=\frac{\pi}{2}.
\end{align*}
Therefore, the lemma follows. \qed

	\begin{thm}\label{thm6}
Let $r \geq \bar{r}_0 >\sqrt[4]{B}$. The geodesics for Eguchi-Hanson metrics with only constant $\theta\in\left( 0, \pi\right) $ and conditions
\begin{align*}
	\lim\limits_{r\rightarrow \bar{r}_0}  t =t_1, \quad	\lim\limits_{r\rightarrow \bar{r} _0}\phi =\phi_1, \quad
		\lim\limits_{r\rightarrow \bar{r} _0}\psi =\psi_1, & \quad
		\lim\limits_{r\rightarrow \bar{r} _0} r^2\frac{d\phi}{dt} =\sqrt{B}\phi_0 ,\\
      \lim\limits_{r\rightarrow \bar{r} _0} \left( \frac{1}{f}\left(\frac{dr}{dt}\right)^{2}
                               +\frac{B\phi_0^2\cos^2\theta}{4r^2f}\right) = & r_1^2-\frac{B\phi_0^2\sin^2\theta}{4\bar{r} _0^2}
\end{align*}		
satisfy
\begin{align*}
		t(r)=&t_1+ \frac{\sqrt{r^4-\left( \frac{B\phi_0^2}{4r_1^2}-r_-\right) r^2-\frac{B^2\phi_0^2\sin^2\theta}{4r_1^2r_-}}}{\varepsilon r_1 \sqrt{r^2-r_-}}\\
	&-\frac{\sqrt[4]{16Br_1^4+\frac{B^2\phi_0^4}{3}}\sqrt{\sin\left( \frac{\eta}{3}+\frac{\pi}{3}\right) }}{\varepsilon \sqrt{2} r_1^2} E\left(\alpha,k^2\right)\\
	&+\frac{\varepsilon \sqrt{2} \bar{r}_0^2}{\sqrt[4]{16Br_1^4+\frac{B^2\phi_0^4}{3}}\sqrt{\sin\left( \frac{\eta}{3}+\frac{\pi}{3}\right) }}F\left(\alpha, k^2\right),\\
	\phi(r) =& \phi_1 +\frac{ \varepsilon\sqrt{2B}\phi_0 }{\sqrt[4]{16Br_1^4+\frac{B^2\phi_0^4}{3}}\sqrt{\sin\left( \frac{\eta}{3}+\frac{\pi}{3}\right) }}F\left(\alpha,k^2\right),\\
		\psi(r) =& \psi_1
			+\frac{ \varepsilon\sqrt{2}B^{\frac{3}{2}}\phi_0\cos\theta\,F(\alpha,k^2)}{(r_-^2-B)\sqrt[4]{16Br_1^4+\frac{B^2\phi_0^4}{3}}\sqrt{\sin\left( \frac{\eta}{3}+\frac{\pi}{3}\right) }} \notag
			\\
			&
			-\frac{ \varepsilon\sqrt{2} B\,\phi_0\,\sqrt[4]{16Br_1^4+\frac{B^2\phi_0^4}{3}}\,\cos\theta \,\sin\frac{\eta}{3}\,\Pi(h_1,\alpha,k^2)}{(r_--\sqrt{B})(\bar{r} _0^2-\sqrt{B})\sqrt{\sin\left( \frac{\eta}{3}+\frac{\pi}{3}\right)}}
			\notag\\
			&+\frac{ \varepsilon\sqrt{2}B\,\phi_0\sqrt[4]{16Br_1^4+\frac{B^2\phi_0^4}{3}}\,\cos\theta \,\sin\frac{\eta}{3}\,\Pi(h_2,\alpha,k^2)}{(r_-+\sqrt{B})(\bar{r} _0^2+\sqrt{B})\sqrt{\sin\left( \frac{\eta}{3}+\frac{\pi}{3}\right)}} ,
		\end{align*}
where
		\begin{align*}
			r_-=&\frac{1}{12r_1^2}\left( B\phi_0^2-2\sqrt{B^2\phi_0^4+48Br_1^4}\cos\left( \frac{\eta}{3}-\frac{2\pi}{3}\right) \right),\\
			\alpha=&\arcsin\sqrt \frac{	12r_1^2r^2-B\phi_0^2+2\sqrt{B^2\phi_0^4+48Br_1^4}\cos\left( \frac{\eta}{3}+\frac{2\pi}{3}\right) }{12r_1^2r^2-B\phi_0^2+2\sqrt{B^2\phi_0^4+48Br_1^4}\cos\left( \frac{\eta}{3}-\frac{2\pi}{3}\right) },\\
            k^2= &\frac{\sin\left( \frac{\eta}{3}+\frac{\pi}{3}\right)}{\sin\left( \frac{\pi}{3}-\frac{\eta}{3}\right)},\\
			h_1=&\frac{B\phi_0^2-2\sqrt{B^2\phi_0^4+48Br_1^4}\cos\left( \frac{\eta}{3}-\frac{2\pi}{3}\right)-12\sqrt{B}r_1^2}{B\phi_0^2-2\sqrt{B^2\phi_0^4+48Br_1^4}\cos\left( \frac{\eta}{3}+\frac{2\pi}{3}\right)-12\sqrt{B}r_1^2} ,  \\
			h_2=&\frac{B\phi_0^2-2\sqrt{B^2\phi_0^4+48Br_1^4}\cos\left( \frac{\eta}{3}-\frac{2\pi}{3}\right)+12\sqrt{B}r_1^2}{B\phi_0^2-2\sqrt{B^2\phi_0^4+48Br_1^4}\cos\left( \frac{\eta}{3}+\frac{2\pi}{3}\right)+12\sqrt{B}r_1^2}.
		\end{align*}			
	\end{thm}
	\pf The geodesic equation (\ref{eq23})  implies that
	\begin{align*}
		\frac{d}{dt}\ln \Big( r^2 \Big| \frac{d\phi}{dt}\Big| \Big)=0.
	\end{align*}
	Thus
	\begin{align}\label{eqf2}
		\frac{d\phi}{dt}=\frac{ \varepsilon\sqrt{B}\phi_0}{r^{2}}.
	\end{align}
	By (\ref{eqf1}) and (\ref{eqf2}), we derive
	\begin{align}\label{eqf3}
		\frac{d\psi}{dt}=\frac{ \varepsilon\sqrt{B}\phi_0\cos\theta(1-f)}{r^{2}f}.
	\end{align}
Substituting
	(\ref{eqf2}) and  (\ref{eqf3}) into (\ref{eq21}), we obtain
	\begin{align*}
		\frac{d}{dt}\left(\frac{1}{f} \left(\frac{dr}{dt}\right)^{2}+ \frac{B\phi_0^2\sin^2\theta}{4r^2}+\frac{B\phi_0^2\cos^2\theta}{4r^2f} \right)=0,
	\end{align*}
	Thus
	\begin{align*}
		\frac{1}{f}\left(\frac{dr}{dt}\right)^{2}+\frac{B\phi_0^2\sin^2\theta}{4r^2}+\frac{B\phi_0^2\cos^2\theta}{4r^2f}=r_1^2.
	\end{align*}
	Therefore
	\begin{align*}
		\frac{d\phi}{dr} &=\frac{d\phi}{dt} \Big/ \frac{dr}{dt}=\frac{\varepsilon \sqrt{B}\phi_0}{r_1r^2}\frac{1}{\sqrt{f-\frac{B\phi_0^2\sin^2\theta}{4r_1^2r^2}f-\frac{B\phi_0^2\cos^2\theta}{4r_1^2r^2}}},\\
		\frac{d\psi}{dr} &=\frac{d\psi}{dt} \Big/ \frac{dr}{dt}=\frac{\varepsilon\sqrt{B}\phi_0 (1-f)\cos\theta}{r_1r^2f}\frac{1}{\sqrt{f-\frac{B\phi_0^2\sin^2\theta}{4r_1^2r^2}f-\frac{B\phi_0^2\cos^2\theta}{4r_1^2r^2}}},\\
			\frac{dt}{dr} &=\frac{\varepsilon }{r_1}\frac{1}{\sqrt{f-\frac{B\phi_0^2\sin^2\theta}{4r_1^2r^2}f-\frac{B\phi_0^2\cos^2\theta}{4r_1^2r^2}}}.
	\end{align*}
	Changing variable $ u=r^2 $,  we obtain
	\begin{align}
		\frac{d\phi}{du} &=\frac{\varepsilon\sqrt{B}\phi_0}{2r_1\sqrt{P(u)}},\label{105}\\
		\frac{d\psi}{du} &=\frac{\varepsilon B^{\frac{3}{2}}\phi_0\cos\theta}{2r_1(u^2-B)}\frac{1}{\sqrt{P(u)}},\label{106}\\
			\frac{dt}{du} &=\frac{\varepsilon u}{2r_1\sqrt{P(u)}},\label{107}
	\end{align}
	with
	\begin{align*}
		P(u)&=u^3-\frac{B\phi_0^2}{4r_1^2}u^2-Bu+\frac{B^2\phi_0^2}{4r_1^2}\sin^2\theta\notag\\
		&=(u-u_2)(u-u_3)(u-u_4).
	\end{align*}

	The cubic equation $ P(u)=0 $ can be denoted in canonical form \cite{FR}
	\begin{equation*}
		w^3+pw+q=0,
	\end{equation*}
	with
	\begin{align*}
		&	p=-B\left( 1+\frac{B\phi_0^4}{48r_1^4}\right) ,\\
		&	q=\frac{B^2\phi_0^2}{4r_1^2}\left(\sin^2\theta-\frac{1}{3}-\frac{B\phi_0^4}{216r_1^4}\right).
	\end{align*}
	Thus the discriminant is
	\begin{align*}
		\Delta=&-(4p^3+27q^2)\\
		=&\frac{B^5\phi_0^8}{256r_1^8}\left( 4\left(\frac{16r_1^4}{B\phi_0^4} \right)^2-(27\sin^4\theta-18\sin^2\theta-1)\frac{16r_1^4}{B\phi_0^4}+4 \sin^2\theta\right).
	\end{align*}
Denote
	\begin{equation*}
		Y=4m^2-(27\sin^4\theta-18\sin^2\theta-1)m+4 \sin^2\theta ,\quad m=\frac{16r_1^4}{B\phi_0^4}>0.
	\end{equation*}
It is obvious that
\begin{align*}
\theta  \neq \frac{\pi}{2} \Lrw Y>0 \Lrw \Delta>0.
\end{align*}
Therefore, the algebraic equation $ P(u)=0 $ has three real roots as follows
	\begin{align*}
		&u_2=\frac{1}{12r_1^2}\left( B\phi_0^2-2\sqrt{B^2\phi_0^4+48Br_1^4}\cos\left( \frac{\eta}{3}+\frac{2\pi}{3}\right) \right) ,\\
		&	u_3=\frac{1}{12r_1^2}\left( B\phi_0^2-2\sqrt{B^2\phi_0^4+48Br_1^4}\cos\left( \frac{\eta}{3}-\frac{2\pi}{3}\right) \right) ,\\
		&u_4=\frac{1}{12r_1^2}\left( B\phi_0^2-2\sqrt{B^2\phi_0^4+48Br_1^4}\cos \frac{\eta}{3}  \right).
	\end{align*}
The theorem follows by integrating (\ref{105}), (\ref{106})  and (\ref{107}) from $ 	\bar{r}_0^2$  to $ r^2 $ respectively. \qed

\begin{thm}
Let constant $r_1$, $\phi_0$ satisfy
\begin{align*}
 r_1>\frac{\sqrt[4]{B}\phi_0 }{2}.
\end{align*}
The geodesics for Eguchi-Hanson metrics constant $\theta \in (0, \pi)$, nonconstant $\phi$ and passing through $\sqrt[4]{B}$ with conditions
	\begin{align*}
		\lim\limits_{r\rightarrow \sqrt[4]{B}}  t =t_1, \quad
		\lim\limits_{r\rightarrow \sqrt[4]{B}} \phi=\phi_1, \quad &
        \lim\limits_{r\rightarrow \sqrt[4]{B}} \psi =\psi_1,\quad
		\lim\limits_{r\rightarrow \sqrt[4]{B}} \frac{d\phi}{dt}=\phi_0, \\
		\lim\limits_{r\rightarrow \sqrt[4]{B}} \frac{1}{\sqrt{f}}\left| \frac{dr}{dt} \right| =&\sqrt{r_1^2-\frac{\sqrt{B}\phi_0^2}{4}}
	\end{align*}
	satisfy
	\begin{align*}
        	t(r)= &t_1+\frac{\varepsilon}{r_1} \Big[\frac{2r_1\sqrt{r^4-B }}{\sqrt{4r_1^2r^2-r_0^4\theta_0^2}}\\
                  &-\sqrt{2}r_0 E\left(\alpha,k^2\right)+\frac{r_0}{\sqrt{2}}F\left(\alpha, k^2\right)\Big],\\
       \theta(r)= &\frac{\pi}{2},\\
		\phi(r) = &\phi_1 +\frac{\varepsilon \sqrt[4]{B}\,\phi_0 }{ \sqrt{2}r_1}F\left(\alpha,k^2\right),\\
        \psi(r) = &\psi_1,\\
	\end{align*}
	where
\begin{align*}
\alpha=\arcsin\sqrt \frac{4r_1^2	r^2-4\sqrt{B}r_1^2}{4r_1^2	r^2-B\phi_0^2 },\quad  k^2= \frac{1}{2}+\frac{\sqrt{B}\phi_0^2}{8r_1^2}.	
\end{align*}
\end{thm}	
\pf Lemma \ref{lemm4-1} implies that $ \theta=\frac{\pi}{2}$. Then (\ref{eqf3}) gives that $\frac{d \psi}{dt}=0$, and it reduces to Theorem \ref{thm3-4}.
\qed

		\mysection{Appendix: Elliptic integrals}\ls\label{5}
		
		In \cite{H}, the incomplete elliptic integrals of the first, the second, and the third kinds are given by
		\begin{align*}
			F(\alpha, k^2)&=\int_{0}^{\alpha}\frac{d\varrho}{\sqrt{1-k^2\sin^2\varrho}},\\
			E(\alpha, k^2)&=\int_{0}^{\alpha}\sqrt{1-k^2\sin^2\varrho}\,d\varrho,\\
			\Pi(h,\alpha, k^2)&=\int_{0}^{\alpha}\frac{d\varrho}{(1-h\sin^2\varrho)\sqrt{1-k^2\sin^2\varrho}},
		\end{align*}
respectively, where $ 0<k<1 $, $ h\in\mathbb{C} $.

The following propositions can be derived straightforwardly.
		
		\begin{prop}
			Let $ a,b,c,x\in\mathrm{R} $, $ x>a>b>c $, and denote
			\begin{equation*}
				f_1(z)=\frac{1}{\sqrt{(z-a)(z-b)(z-c)}},
			\end{equation*}
			then
			\begin{equation*}
				\int_{a}^{x}f_1(z)dz=\frac{2}{\sqrt{a-c}}F\left(\alpha,k^2\right),
			\end{equation*}
			where
\begin{align*}
\alpha= \arcsin\sqrt{\frac{x-a}{x-b}}, \quad k^2= \frac{b-c}{a-c}.
\end{align*}
		\end{prop}
		
		\begin{prop}
			Let $ a,b,c,x\in\mathrm{R} $, $ x>a>b>c $, and denote
			\begin{equation*}
				f_2(z)=\frac{z}{\sqrt{(z-a)(z-b)(z-c)}},
			\end{equation*}
			then
			\begin{align*}
				\int_{a}^{x}f_2(z)dz=&\frac{2\sqrt{(x-a)(x-c)}}{\sqrt{x-b}}-2\sqrt{a-c}E\left(\alpha,k^2\right)\\
				 & +\frac{2a}{\sqrt{a-c}}F\left(\alpha,k^2\right),
			\end{align*}
			where
\begin{align*}
\alpha= \arcsin\sqrt{\frac{x-a}{x-b}}, \quad k^2= \frac{b-c}{a-c}.
\end{align*}
		\end{prop}
		
		\begin{prop}
			Let $ a,b,c,d, x\in\mathrm{R} $, $ x>a>b>c $, $ a>d $, and denote
			\begin{equation*}
				f_3(z)=\frac{1}{(z^2-d^2)\sqrt{(z-a)(z-b)(z-c)}},
			\end{equation*}
			then
			\begin{align*}
				\int_{a}^{x}f_3(z)dz
				=&\frac{2F(\alpha,k^2)}{(b^2-d^2)\sqrt{a-c}}-\frac{(a-b)\Pi(h_1,\alpha,k^2)}{d(b-d)(a-d)\sqrt{a-c}}\\
                 &+\frac{(a-b)\Pi(h_2,\alpha,k^2)}{d(b+d)(a+d)\sqrt{a-c}},
			\end{align*}
			where
\begin{align*}
\alpha= \arcsin\sqrt{\frac{x-a}{x-b}}, \quad k^2= \frac{b-c}{a-c}, \quad h_1=\frac{b-d}{a-d}, \quad h_2=\frac{b+d}{a+d}.
\end{align*}
		\end{prop}
		
		\begin{prop}
			Let $ a_1, a_2, a_3, a_4, a_5, x\in\mathrm{R} $, $ x>a_1>a_2>a_3$, $ a_1>a_4 \geq a_5$,  and denote
			\begin{equation*}
				f_4(z)=\frac{z^2}{(z-a_4)(z-a_5)\sqrt{(z-a_1)(z-a_2)(z-a_3)}},
			\end{equation*}
			then
			\begin{align*}
				\int_{a}^{x}f_4(z)dz
				=&\frac{2a_2^2}{(a_2-a_4)(a_2-a_5)\sqrt{a_1-a_3}}F(\alpha,k^2)\notag\\
				&-\frac{2a_4^2(a_1-a_2)}{(a_4-a_5)(a_1-a_4)(a_2-a_4)\sqrt{a_1-a_3}}\Pi(h_1,\alpha,k^2)\notag\\
				&+\frac{2a_5^2(a_1-a_2)}{(a_4-a_5)(a_1-a_5)(a_2-a_5)\sqrt{a_1-a_3}}\Pi(h_2,\alpha,k^2),
			\end{align*}
			where
\begin{align*}
\alpha= \arcsin\sqrt{\frac{x-a_1}{x-a_2}}, \quad k^2= \frac{a_2-a_3}{a_1-a_3}, \quad h_1=\frac{a_2-a_4}{a_1-a_4}, \quad h_2=\frac{a_2-a_5}{a_1-a_5}.
\end{align*}
		\end{prop}

{\footnotesize {\it Acknowledgement. This work is supported by the special foundation for Guangxi Ba Gui Scholars and Junwu Scholars of Guangxi University.}


\begin{thebibliography}{99}
			
			\bibitem{BE} E. Battista, G. Esposito, Geodesic motion in Euclidean Schwarzschild geometry, Eur. Phys. J. C 82 (2022) 1088.
			\bibitem{C} S. Chandrasekhar, The mathematical theory of black holes, Oxford University Press, New York, 1983.
            \bibitem{CM} A. Cie\'{s}lik, P. Mach, Revisiting timelike and null geodesics in the Schwarzschild spacetime: general expressions in terms of Weierstrass elliptic functions, Class. Quantum Grav. 39 (2022) 225003.
			\bibitem{EH1} T. Eguchi, A.J. Hanson, Asymptotically flat self-dual solutions to Euclidean gravity, Phys. Lett. 74B (1978) 249-251.
			\bibitem{EH2} T. Eguchi, A.J. Hanson, Self-dual solutions to Euclidean gravity,  Ann. Phys. 120 (1979) 82-106.
			\bibitem{FR} R. Franci, L.T. Rigatelli, Storia della Teoria delle Equazioni Algebriche, Mursia, Milano,1979.
			\bibitem{GH} G.W. Gibbons, S.W. Hawking, Classification of gravitational instanton symmetries, Commun. Math. Phys.  66 (1979) 291-310.
			\bibitem{H} H. Hancock, Elliptic Integral, Nabe Press, Carolina, 2010.
            \bibitem{HP} S.W. Hawking, C.N. Pope, Symmetry breaking by instantons in supergravity, Nuclear Phys. B 146 (1978) 381-392.
            \bibitem{L} C. LeBrun, Counter-examples to the generalized positive action conjecture, Commun. Math. Phys. 118 (1988) 591-596.
			\bibitem{Z} X. Zhang, Scalar flat metrics of Eguchi-Hanson type, Commun. Theor. Phys.  42 (2004) 235-237.
			
		\end{thebibliography}
	\end{document}